# Non-monotonic zero point entropy in diluted spin ice

X. Ke[1], R. S. Freitas[1*], B. G. Ueland[1†], G. C. Lau[2], M. L. Dahlberg[1], R. J. Cava[2], R. Moessner[3], and P. Schiffer[1]

1. Department of Physics and Materials Research Institute, Pennsylvania State University, University Park, PA, 16802.

2. Department of Chemistry and Princeton Materials Institute, Princeton University, NJ, 08540.

3. Rudolf Peierls Centre for Theoretical Physics, Oxford University, 1 Keble Road, Oxford OX1 3NP, UK.

Water ice and spin ice are important model systems in which theory can directly account for "zero point" entropy associated with quenched configurational disorder. Spin ice differs from water ice in the important respect that its fundamental constituents, the spins of the magnetic ions, can be removed through replacement with non-magnetic ions while keeping the lattice structure intact. In order to investigate the interplay of frustrated interactions and quenched disorder, we have performed systematic heat capacity measurements on spin ice materials which have been thus diluted up to 90%. Investigations of both Ho and Dy spin ices reveal that the zero point entropy depends non-monotonically on dilution and approaches the value of Rln2 in the limit of high dilution. The data are in good agreement with a generalization of Pauling's theory for the entropy of ice.

Entropy is one of the most important concepts in thermodynamics, but it is rarely directly connected through experiment to its statistical definition rooted in the number of available states. One such connection is found in the case of water ice wherein each oxygen atom is bonded to four hydrogen atoms. The lowest energy state has two hydrogens located closer to the oxygen with the other two further away, establishing the so-called 'ice rules' [1]. This arrangement allows for a high degeneracy of states, and a resultant 'zero point entropy' has been measured thermodynamically and can be associated with the disordered states frozen in place as $T \rightarrow 0$ [2].

A high degeneracy of states also leads to a range of exotic physics in geometrically frustrated magnets, including 'spin ice' materials [3, 4, 5, 6] in which the low temperature behavior is closely analogous to that of water ice. In these materials, the magnetic ions ($Ho^{3+}$ or $Dy^{3+}$) occupy a pyrochlore lattice of corner-sharing tetrahedra, and the local crystal field environment causes the magnetic moments to point along the lines connecting the centers of two tetrahedra at low temperatures [5, 6]. This strong single-ion Ising anisotropy, combined with dipolar and exchange interactions [7], results in a highly degenerate two-in/two-out spin configuration for the ground states of these materials [3, 5] – locally equivalent to the situation for hydrogen atoms in water ice. Although numerical calculations [8] suggest the existence of a long range ordered state for these systems which would have no zero point entropy, the spins in these materials freeze into a non-equilibrium low temperature state with approximately the same zero point entropy as water ice [4, 9], and no long range ordering has been observed experimentally in zero magnetic field.

Here we examine experimentally how dilution of the spin ice lattice affects the zero point entropy over a broad range of dilution in both Ho and Dy spin ice materials -- a measurement which would not be possible in water ice. We find that the zero point entropy of diluted spin ice depends *non-monotonically* on the level of dilution, with the zero point entropy per spin approaching the expected full Rln2 in the limit of high dilution. We explain these results by an extension of Pauling's theory for the zero point entropy of ice.

Previous workers have examined the effects of chemical doping on the entropy of water ice. Heat capacity measurements on HF-doped ice crystals show a decrease of a few percent in the zero point entropy, an effect which is attributed to the accelerated configurational ordering of protons [10]. Doping of KOH into the ice crystal leads to the formation of a long range ordered state and an elimination of most of the zero point entropy [11]. The spin ice system, by contrast, allows a much broader range of doping and the introduction of disorder. For example, "stuffing" extra magnetic ions into sites of the non-magnetic cations [12] does not appear to change the zero point entropy per spin for stuffed $Ho_2(Ti_{2-x}Ho_x)O_{7-x/2}$. An alternative approach, which we pursue here, is to replace a fraction of the magnetic ions in spin ice with non-magnetic ions such as Y, Lu, or La, a substitution which has been used previously to examine the spin dynamics of these materials [13, 14, 15, 16].

Polycrystalline samples of $Dy_{2-x}Lu_xTi_2O_7$, $Dy_{2-x}Y_xTi_2O_7$ and $Ho_{2-x}Y_xTi_2O_7$ were prepared using standard solid-state synthesis techniques. X-ray diffraction confirmed all samples to be single phase with the pyrochlore structure. The saturation magnetization at 1.8 K of all samples in a magnetic field of 5.5 T (measured with a Quantum Design

MPMS) was 5.0 $\mu_B$ per rare earth ion to within 5%. This is about half the expected full moment for both Dy and Ho ions obtained from temperature dependent low field DC susceptibility measurements, and it indicates that the rare earth moments retain their Ising-like nature in the diluted samples. We also performed ac susceptibility measurements with the ACMS option of the Quantum Design PPMS and specific heat measurements with a PPMS cryostat with the $^3$He option, using a standard semiadiabatic heat pulse technique. The Dy-based samples were pressed with Ag to facilitate thermal equilibration and the Ho-based samples were pressed directly to form pellets.

In Fig. 1, we show the real part of the ac susceptibility, $\chi'$, as a function of temperature for diluted $Dy_{2-x}Y_xTi_2O_7$ samples with $x = 0, 0.4, 1.4, 1.8$ at a frequency of 1 kHz. As has been previously observed [13, 14, 17], there are two frequency dependent maxima appearing in $\chi'(T)$, characterizing the dynamic spin freezing in these materials. The maxima in $\chi'(T)$ indicate that the spin relaxation time is longer than the experimental measurement time and that the spins are falling out of equilibrium below that temperature. The high temperature peak is associated with a single spin process [14, 15], and the re-entrance of the spin freezing transition at higher dilution is attributed to a combination of quantum mechanical and thermal processes combined with changes in the crystal field level spacing [14]. The low temperature peak in $\chi'(T)$ originates from the spin freezing associated with the spins falling into compliance with the ice rules and freezing into a disordered state.

In order to measure the zero point entropy of the diluted spin ice samples, we measured the specific heat, $C(T)$. We subtracted the addendum heat capacity associated with the sample holder, the phonon contributions (from a fit for the Debye coefficient), and the hyperfine contributions (in the case of the Ho materials). We were thus able to obtain the magnetic specific heat associated with the Ising-like rare earth spins, $C_{mag}(T)$. We then integrated $C_{mag}(T)/T$ from the lowest temperatures to obtain the measurable magnetic entropy of the system, $S_{meas}$, at temperatures well above the spin freezing transitions shown in figure 1. Applying this method to an ideal Ising system with zero entropy in the low temperature limit would result in an integrated measurable molar entropy of $S_{meas} = $ Rln2 corresponding to the full $2^N$ states available to $N$ spins (R is the gas constant). An integrated molar entropy of less than Rln2 indicates the presence of entropy at the lowest temperatures studied. The difference between the integrated entropy and Rln2 represents a measurement of the zero point entropy.

Fig. 2 shows $C_{mag}(T)$ for $Dy_{2-x}Y_xTi_2O_7$ with $x = 0, 0.4, 1.4, 1.8$ at $H = 0$ T (a) and $H = 1$ T (b). Unlike KOH-doped water ice [11], no long range ordering is observed for any diluted samples down to $T = 0.4$ K (our lowest temperature), a result which is also implied by the broad peaks in $C_{mag}(T)$. Furthermore, the zero field specific heat is strongly suppressed for larger $x$, suggesting that the zero point entropy per spin in the low temperature limit is larger. This difference is suppressed in a field of $H = 1$ T, where the data from different samples are virtually identical.

In Fig. 3, we plot the magnetic entropy, $S_{meas}(T)$, with $H = 0$ T (a) and $H = 1$ T (b). For undiluted $Dy_2Ti_2O_7$ at zero magnetic field, the entropy at high temperature is around 4.07 J/(K•mol$_{Dy}$), consistent with previous results [4]. This is smaller than Rln2 by almost exactly the expected zero point entropy ½Rln(3/2). The integrated entropy for the diluted samples is smaller than that for undiluted $Dy_2Ti_2O_7$, as shown in Fig. 3(a), indicating an increase in the zero point entropy. Applying a magnetic field to lift the degeneracy of the ground states leads to an increased integrated measured entropy (i.e., a reduction of the zero point entropy) for all of the samples, as has been previously reported for undiluted $Dy_2Ti_2O_7$ [4].

The magnetic entropy integrated up to 16 K at zero applied field is shown as a function of $x$ in Fig. 4. Remarkably, the magnetic entropy does not depend monotonically on dilution. As seen clearly for diluted $Dy_{2-x}Y_xTi_2O_7$ at zero field, the measured entropy decreases slightly upon dilution, then increases slightly upon further dilution with a maximum around $x = 1$, before sharply dropping with further dilution until it approaches zero in the limit of full dilution. Similar behavior is also observed in $Ho_{2-x}Y_xTi_2O_7$ samples, and the entropy of $Dy_{1.0}Lu_{1.0}Ti_2O_7$ is consistent with the Y doped samples (data not shown). This non-monotonic dependence of magnetic entropy on dilution is removed in a 1 T magnetic field, presumably because of the lifted degeneracy of the ground states due to a large Zeeman interaction.

We estimated our uncertainty in the entropy measurement by repeating the measurement using different samples for $x = 0, 0.2, 1.8$. The difference in the resulting entropy is small, with the maximum being about 0.12 J/(K•mol$_{Dy}$) for $H = 0$ T, as indicated by the error bar in Fig. 4. Some uncertainty is also introduced by our inability to measure the heat capacity all the way down in temperature to $T = 0$, a problem which is encountered by every entropy measurement of this sort. A low temperature peak in $C_{mag}(T)$ associated with spin ordering [8] would eliminate the zero point entropy, but the spins in undiluted spin ice fall out of equilibrium at sufficiently high temperatures that such ordering does not occur [3, 4]. Low temperature susceptibility data taken in a dilution refrigerator on a $x = 1.8$ sample, shown in the inset to Fig. 2(a), indicate that this is also the case for diluted samples, and therefore suggest that little additional entropy would be obtained by integrating the heat capacity from lower temperatures. (Linear extrapolations of $C_{mag}(T)$ to absolute zero temperature suggests that this contribution to the entropy would be at most on the order of 15% and that it would not change the non-monotonic behavior of the entropy.)

The observed non-monotonic variation in the measured entropy at zero field has its origin at low temperature where interactions are strong: note that in Fig. 2(a), the $C_{mag}(T)$ curves with $x = 0.4$ and $x = 1.4$ cross at $T \sim 2$ K. At higher temperature, our data are consistent with a monotonic dependence of $S_{meas}$ on $x$, in accordance with the asymptotic result for the entropy of an Ising model, which only depends on the number of interaction partners and not on further geometric details [18]. To model the zero point entropy, we thus need a 'non-perturbative' approach at low temperature, which takes into account the geometrical frustration present in spin ice. Such an approach is provided by the Pauling estimate for the entropy, which, for the case of the undiluted system, is known to be quite accurate [4]. Here, we generalize it to the diluted case.

The basic idea of the Pauling approximation is that the ground state constraints of each tetrahedron are treated as independent. The resulting formula for the entropy as a function of $x$ is given by

$$S_{meas}(x)/R = \ln 2 - (\ln W(x/2))/[N(1-x/2)]$$

where $W(x/2) = 2^{N(1-x/2)} \prod_{i=1}^{4} f_i^{n_i(x/2)}$ and $x/2$ is the ratio of the non-magnetic dilution. Here, $N$ is the total number of sites of the lattice (occupied or not). And $n_i(x/2)$ denotes the number of tetrahedra with $i$ sites occupied by spins:

$$n_i(x/2) = \tfrac{N}{2} C_4^i (1-x/2)^i (x/2)^{(4-i)}$$

$f_i$ is the crucial element of the Pauling estimate: it denotes what fraction of configurations of a tetrahedron with $i$ sites occupied by spins respect the ground state condition. For $i = 4$, there are 6 out of the 16 possible configurations in the ground state, namely those in which two spins point into the tetrahedron, and two out: $f_4 = 3/8$. If one spin is missing, there are still 6 ground states (2-in 1-out or 1-in 2-out) but only 8 possible configurations: $f_3 = 3/4$. Similarly, $f_2 = 1/2$, and $f_1 = 1$.

In Fig. 4, we plot the Pauling estimate for the entropy of spin ice. Most saliently, it does reproduce the non-monotonic behavior of $S_{meas}(x)$, first decreasing, then increasing, and finally decreasing again. This is understood simply as a consequence of the non-monotonic variation of the $f_i$: for weak dilution, a few tetrahedra with only three spins replace fully occupied ones. As $f_3 > f_4$, the presence of such tetrahedra is less constraining, and $S_{meas}(x)$ decreases. However, upon diluting more strongly, tetrahedra with only two spins start appearing, and as their number grows at the expense of three-spin tetrahedra, $f_2 < f_3$ leads to an increase in $S_{meas}(x)$. Finally, since tetrahedra containing only one spin are always in a ground state ($f_1 = 1$), in the limit of strong dilution $S_{meas}$ attains its minimum possible value, $S_{meas}(x \to 2) = 0$ [19]. This is consistent with expectations for free spins, which have zero point entropy of Rln2 (i.e., an absence of any correlations).

As shown in Fig. 4, the Pauling estimates agree strikingly well with the experimental data on the diluted spin ice samples. The agreement is good in that the general features of the curves are captured by the approximation. Indeed, the difference between the theoretical curve and the experiments is as big as the difference between the measurements on Dy and Ho spin ice. This indicates that an improvement in the agreement will require consideration of compound-specific features. Perhaps the one systematic difference between theory and experiment is that the non-monotonicity in the residual entropy appears more pronounced in the latter; this might indicate correlations between the locations of the Y ions.

These results point to the rich physics available in ice-like systems for studying the fundamental physics of entropy, and they represent a well-controlled model for the study of the interplay of disorder and interactions. While we have explored one method of introducing disorder to the spin ice system, further studies could also probe the introduction of different magnetic species or the introduction of small amounts of lattice disorder by making substitutions for the Ti ions. A further interesting avenue of exploration would be the comparison of these diluted systems' entropy with that in spin glass systems of various density (for instance, $SrGa_{12-x}Cr_xO_{19}$ [20]) or the statistical entropy of the recently developed "artificial" frustrated magnets [21].

We acknowledge the financial support from NSF grant DMR-0353610 and R.S.F. thanks CNPq-Brazil for sponsorship.

FIGURE CAPTIONS

Figure 1: The real part of ac susceptibility, $\chi'$, as a function of temperature for diluted $Dy_{2-x}Y_xTi_2O_7$ samples with $x$ = 0, 0.4, 1.4, 1.8 at a characteristic frequency of 1 kHz and zero field.

Figure 2: Magnetic specific heat $C_{mag}$ as a function of temperature for diluted $Dy_{2-x}Y_xTi_2O_7$ samples with $x$ = 0, 0.4, 1.4, 1.8 at $H$ = 0 T (a) and $H$ = 1 T (b). The inset shows the low temperature dependence of the real part of the $f$ = 50 Hz ac susceptibility, $\chi'$, measured in a dilution fridge.

Figure 3: The integrated magnetic entropy as a function of temperature for diluted $Dy_{2-x}Y_xTi_2O_7$ samples with $x$ = 0, 0.4, 1.4, 1.8 at $H$ = 0 T (a) and $H$ = 1 T (b).

Figure 4: The entropy at $T$ = 16 K as a function of dilution $x$ for $Dy_{2-x}Y_xTi_2O_7$ and $Ho_{2-x}Y_xTi_2O_7$ samples at $H$ = 0 T. The dashed purple curve represents the theoretical calculation based on Pauling estimate.

Figure 1.
X. Ke, et al.

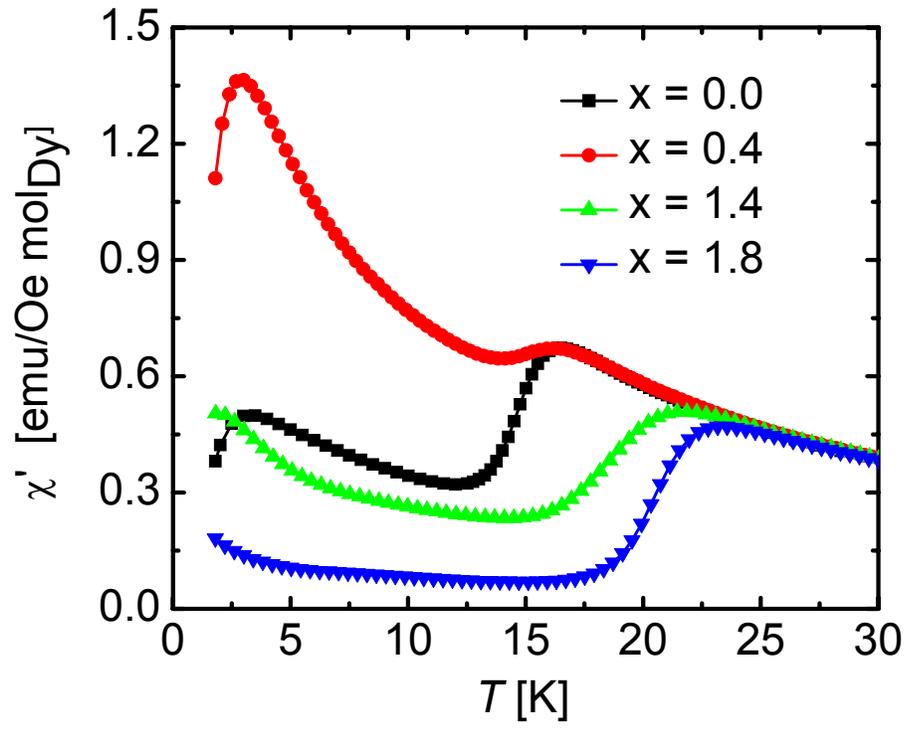

Figure 2.
X. Ke, et al.

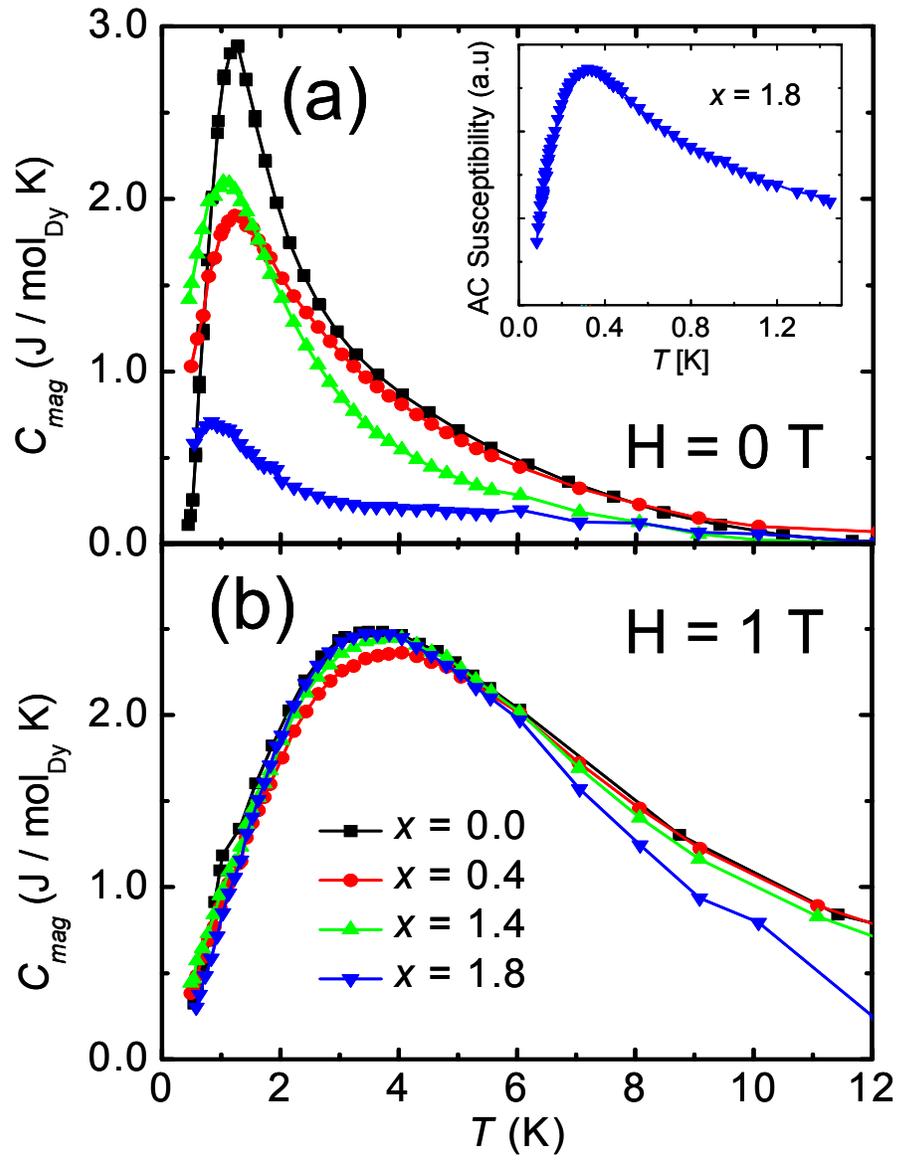

Figure 3.
X. Ke, et al.

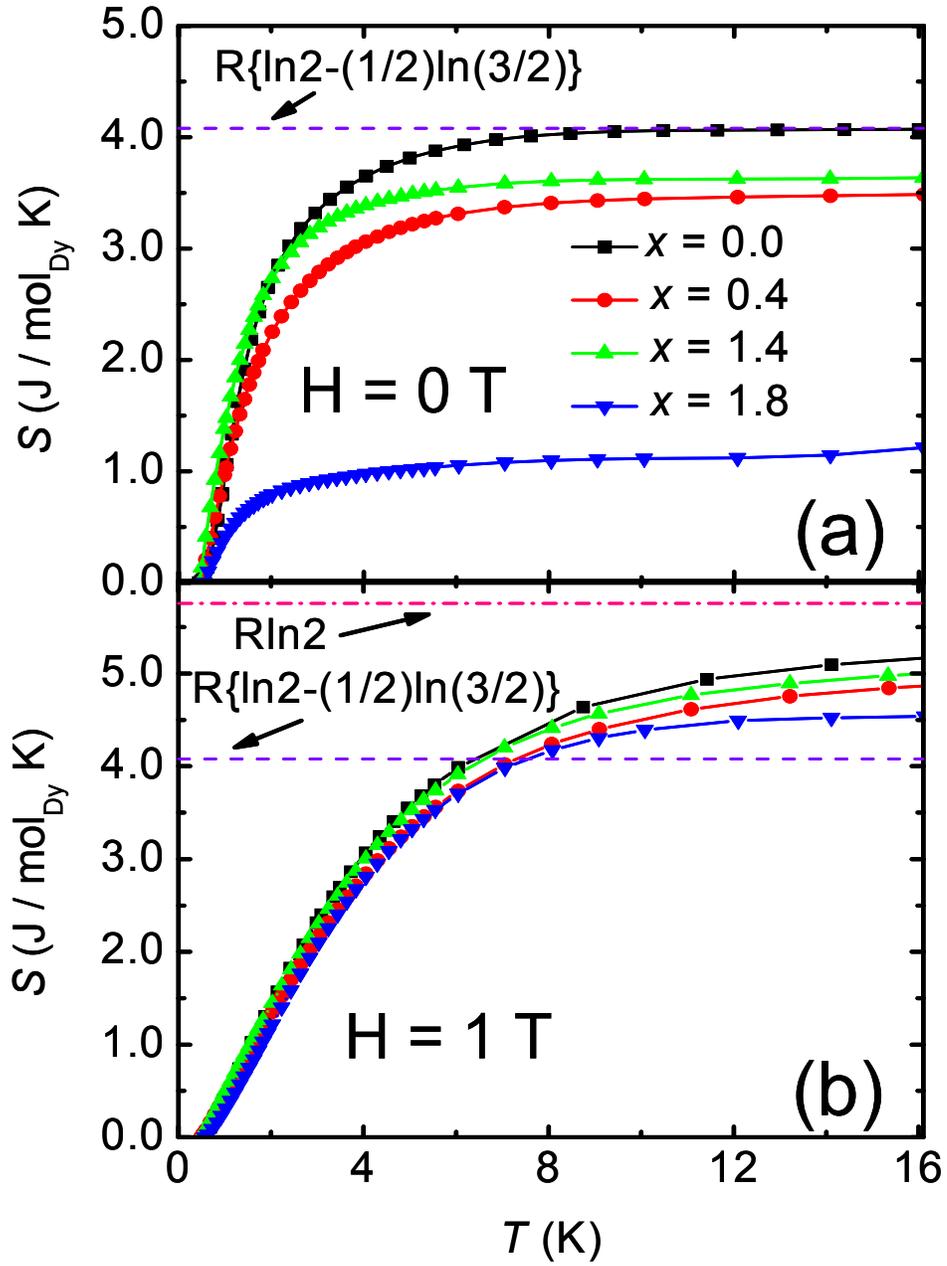

Figure 4.
X. Ke, et al.

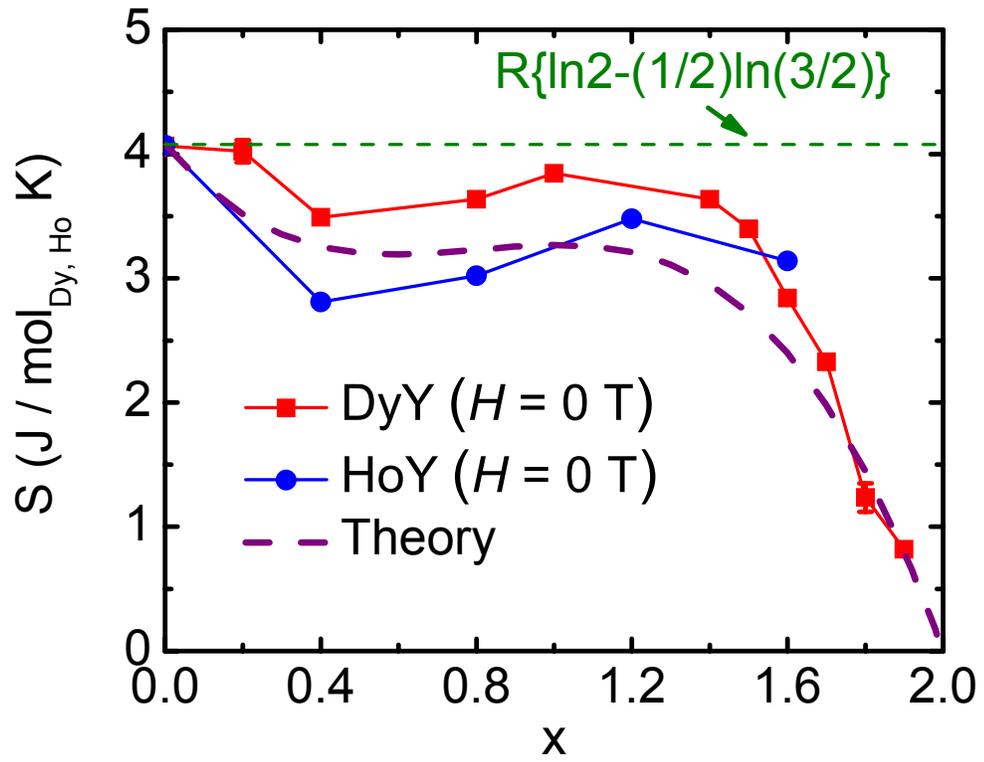

* Present address: Instituto de Física, Universidade de São Paulo, Caixa Postal 66318, 05315-970, São Paulo, SP, Brazil.

† Present address: National Institute of Standards and Technology, NIST Center for Neutron Research, Gaithersburg MD 20899.